# An accurate moving wall boundary algorithm for Direct Simulation of Monte Carlo in unsteady rarefied flow

**He Zhang (张贺), Fanli Shan (单繁立),[a) ]Hong Fang (方洪), Xing Zhang (张星), Jun Zhang (张军), Jinghua Sun (孙精华)**

## AFFILIATIONS

Science and Technology on Space Physics Laboratory, Beijing 100076, P.R. China

[a)] **Author to whom correspondence should be addressed**: tacyon@163.com

## ABSTRACT

An accurate algorithm is proposed to improve the prediction of a particle in collision with a moving wall within the direct simulation Monte Carlo (DSMC) framework for the simulation of unsteady rarefied flows. This algorithm is able to predict the particle-wall collision in a coupled manner by removing the assumption employed by the approximate algorithm, in that the wall is frozen during the collision. The trajectory equation of the particle is theoretically constructed in a moving object coordinate system. It can accurately describe the geometries of the collision between a particle and an arbitrarily shaped object of which the motion incorporates both translation and rotation, thus allowing to deal with complex problems. In contrast, the approximate algorithm ignores the effect of the moving wall on the particle movement during the collision, and therefore induces error that is an increasing function of the wall velocity. Four rarefied flow problems are applied to validate the accurate algorithm. It is shown that the algorithm can produce results perfectly consistent with the Maxwellian theoretical solutions and ensure particle conservation to avoid gas leakage. It is also shown in a three-dimensional case of a re-entry module that the steady simulation fails to reproduce the hysteresis effect while the unsteady simulation using the accurate algorithm can do that, indicating that the unsteady simulation with an appropriate algorithm as proposed in the present work is essentially required in such applications.

## I. INTRODUCTION

When the molecular mean free path ( $\lambda$ ) is comparable to or larger than the flow characteristic length scale ( $L$ ), insufficient molecular collisions will result in the gas distribution function not being maintained at the near-equilibrium state. The Knudsen number, defined as $Kn = \lambda/L$ , is used to describe the deviation from the near-equilibrium state. For high Knudsen number flows, the strongly non-equilibrium state will affect gas transport properties, such as the viscosity and heat conduction, which results in the failure of the Navier-Stokes (NS) equations. Both rarefied and micro flows correspond to high Knudsen number. Therefore, the low orbit satellite,[1,2] space vehicle,[3,4] vacuum equipment[5] and micro-electro-mechanical systems (MEMS),[6,7] etc., in which rarefied or micro flow phenomena exist, are posed with problems of high Knudsen number flows.

Among the methods for the simulations of high Knudsen number flows,[8-16] the direct simulation Monte Carlo (DSMC)[16,17] is the most widely used. This method has been successfully applied in a variety of studies involving rarefied or micro flows.[2,3,7,18-24] Unlike the methods of solving partial differential equations, the DSMC method employs the Monte Carlo randomized trials to predict molecular behaviors. However, even for rarefied flows, the number of molecules is so huge that it is unaffordable to track each of them when conducting the simulation. For this reason, two simplifications are introduced to this method, aimed at reducing the amount of computation to an acceptable level. The first simplification is that each DSMC particle is regarded as a representative of a large number of molecules. The number of molecules represented by a DSMC particle is termed the scaling factor.[25] The second simplification assumes that particle movement and collision are decoupled, in that a particle first moves in a straight line and then collides with another. During the collision, the collision selection scheme is introduced to stochastically select two particles to form a collision pair. Among different collision selection schemes, the no time counter (NTC) scheme, proposed by Bird,[26] is most popular and widely used due to its simplicity and high efficiency.

In the DSMC method, the collision between particle and wall is described by the gas-surface interaction model. The most commonly applied gas-surface interaction model is the Maxwell model.[16] This model is highly simplified since it neglects complex scattering characteristics of particles, and therefore, it has an advantage of easy implementation in the simulation. Other models, such as the Cercignani-Lampis-Lord (CLL) model,[27] introduce additional parameters to indicate scattering characteristics, which makes them more physically realistic in comparison with the Maxwell model, but at a cost that limits their widespread use for lack of knowledge of appropriate values of the parameters.

The DSMC method was usually used to simulate steady rarefied flows in prior works. However, there are unsteady rarefied flows in many engineering applications, such as the separations of spacecrafts and the movements of mechanisms in MEMS. Although the DSMC is essentially a time-advancing method, its original procedure is not quite applicable for unsteady simulations. For instance, the NTC scheme sets the collision rate as proportional to $N\bar{N}/V$ with $N$ and $\bar{N}$ representing the instantaneous and temporally averaged numbers of particles in a cell whose volume is $V$, respectively. It is inappropriate to employ $\bar{N}$ to compute the collision rate in unsteady simulations due to the fact that the particle number varies with time. For this reason, a previous research[28] suggests to apply $N(N-1)/V$ [29] instead of $N\bar{N}/V$ for the collision rate computation. In addition to this, an unsteady simulation requires an unsteady sampling to smooth fluctuations, which can be achieved by an ensemble average or an increase of particles but at a cost of a massive computation. To make the computational expenditure acceptable, Bird's DS2V code applies an approach that averages a number of time steps over an interval centered on the sampling time to achieve the unsteady sampling. This approach, however, gives rise to a problem in that the results are "smeared". An improved approach, termed the temporal variable time step (TVTS),[30] is proposed by Cave et al. to deal with this problem. It can weaken the smearing to the results by reducing the time step before and after the sampling time.

Another challenge for using the DSMC method in unsteady simulations is how to construct an algorithm that is capable of dealing with moving walls. Bird[16] proposed a one-dimensional piston boundary algorithm and incorporated it into the DSMC method to predict shock wave formation and reflection induced by piston movement. This algorithm can be extended to two- and three-dimensional simulations[31-34] only in the case that the motion of wall is translational. An approximate algorithm within the DSMC framework, which assumes that the wall is stationary during the particle-wall collision, thus avoiding complex geometric computation, has also been applied for dealing with moving walls.[31] As presented in section III, the error accumulation rate induced by this algorithm is increased with the normal velocity of wall. Therefore, it is appropriate to apply this algorithm in MEMS applications in which the wall moves slowly,[31] but it is potentially inappropriate in spacecraft applications in which the wall may move at a high speed that tends to induce a nonnegligible error accumulation. As a result, an algorithm, which is capable of accurately dealing with moving walls, is essentially required for the DSMC simulation. It is noted that the studies on this topic are rarely seen in open literatures. Although the complex motions incorporating translation and rotation of the wall were involved in some prior studies,[35-37] the mathematical details of the algorithms used were not provided. For this reason, an accurate algorithm is proposed and detailed in the present work. This algorithm is constructed for arbitrarily shaped object of which the motion incorporates both translation and rotation. The use of the algorithm allows an accurately geometric description of the particle-wall collision by acquiring the collision position from solving the particle trajectory equation in a moving object coordinate system. Thus, it essentially improves predictions of unsteady rarefied flows.

The following sections first give the stationary wall boundary condition and the approximate algorithm as well as the error analysis of this algorithm, then detail the construction of the proposed accurate algorithm, and finally validate it in numerical studies.

## II. STATIONARY WALL BOUNDARY CONDITION

Because the DSMC method assumes that particle movement and collision are decoupled, the position of a particle after a straight-line motion within a time step is expressed as below.

$$\mathbf{P}_1 = \mathbf{P}_0 + \Delta t \mathbf{V}_0 \tag{1}$$

Here, $\Delta t$ is the time step, $\mathbf{P}_0$ and $\mathbf{P}_1$ represent the particle position vectors at the moments $t_0$ and $t_1$ ($t_1 = t_0 + \Delta t$), respectively, and $\mathbf{V}_0$ denotes the particle velocity vector at $t_0$. In the case that the line segment $\mathbf{P}_0\mathbf{P}_1$ has a geometric intersection with a stationary wall, the particle will collide with the wall and the collision moment $t_c$ is determined as follows.

$$t_c = t_0 + \frac{\left|\mathbf{P}_c - \mathbf{P}_0\right|}{\left|\mathbf{P}_1 - \mathbf{P}_0\right|} \Delta t \tag{2}$$

Here, $\mathbf{P}_c$ represents the collision position vector and it can be determined according to the geometric intersection between the wall and particle trajectory.

In the DSMC method, the particle reflection velocity from a wall has to be modeled. The Maxwell model, which was first used as the wall boundary condition of the Boltzmann equation, has been widely applied for determining this velocity. This model assumes that the reflection of a particle from a wall is either specular or diffusive and adopts the half Maxwellian velocity distribution[16] as presented in Eq. (3) to describe the diffusive case ( $f_{dif}$ ).

$$f_{dif}\left(\mathbf{V}_r\right) = 2\left(\frac{m}{2\pi k_B T_w}\right)^{\frac{3}{2}} \exp\left(-\frac{m\mathbf{V}_r^2}{2KT_w}\right) \tag{3}$$

Here, $T_w$ is the wall temperature, $m$ and $\mathbf{V}_r$ denotes the particle mass and the reflection velocity vector, respectively, and $k_B$ stands for the Boltzmann constant. By presuming that the specular reflection velocity follows the Dirac delta distribution, the overall reflection velocity distribution $f_r$ can be obtained according to the weighted average algorithm as below.

$$f_r\left(\mathbf{V}_r\right) = \left(1 - \sigma_\tau\right)\delta\left(\mathbf{V}_r - \mathbf{V}_{spe}\right) + \sigma_\tau f_{dif}\left(\mathbf{V}_r\right) \tag{4}$$

Here, $\sigma_\tau$ is the wall accommodation coefficient indicating the proportion of the diffusive reflection, $\delta$ represents the Dirac delta distribution, and $\mathbf{V}_{spe}$ denotes the specular reflection velocity vector that is determined by the incident velocity vector. After $\mathbf{V}_r$ is obtained from Eq. (4) with a randomized sampling, $\mathbf{P}_1$, which incorporates the effect of the particle-wall collision, can be acquired from the following equation.

$$\mathbf{P}_1 = \mathbf{P}_c + \mathbf{V}_r\left(t_1 - t_c\right) \tag{5}$$

## III. APPROXIMATE MOVING WALL BOUNDARY ALGORITHM

The boundary condition presented in section II is only applicable to the case of stationary wall, and it must be modified when applied to the case of moving wall. Firstly, a moving wall indicates that the particle incident velocity in the Maxwell model is no longer the particle velocity itself, but the velocity relative to the wall as shown in Eq. (6).

$$\begin{aligned} \mathbf{V}_{i,\,\mathrm{Maxwell}} &= \mathbf{V}_0 - \mathbf{V}_w \\ \mathbf{V}_r &= \mathbf{V}_{r,\,\mathrm{Maxwell}} + \mathbf{V}_w \end{aligned} \tag{6}$$

Here, $\mathbf{V}_w$ denotes the wall velocity vector; $\mathbf{V}_{i,\,\mathrm{Maxwell}}$ and $\mathbf{V}_{r,\,\mathrm{Maxwell}}$ denotes respectively the incident velocity vector required by the Maxwell model and the reflection velocity vector obtained from this model. Secondly, it is more challenging to determine $\mathbf{P}_c$ compared to the case of stationary wall. Figure 1 illustrates an algorithm to address this issue. It freezes the wall during the particle-wall collision, and moves the wall to its new position after $\mathbf{P}_1$ is determined. Although it is easy to implement, it is essentially an approximate algorithm and thus induces error. The analytical evaluation of the error is given in the following.

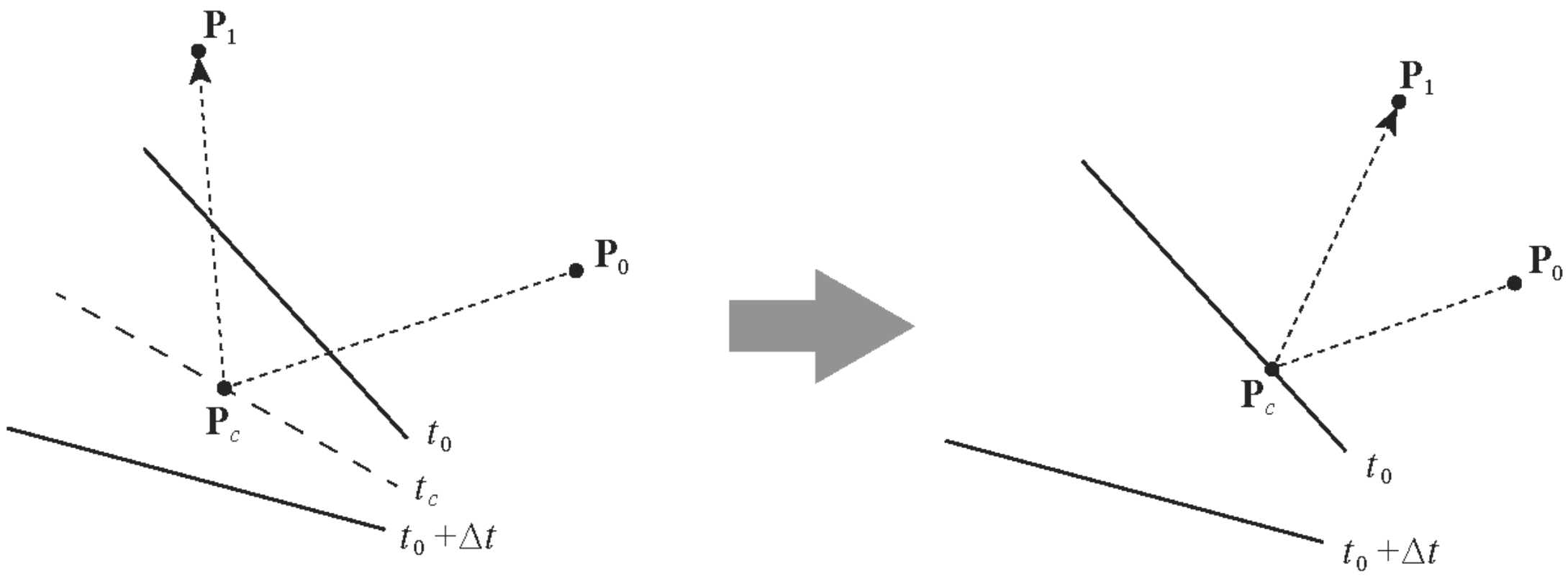

**FIG. 1.** Approximate moving wall boundary algorithm.

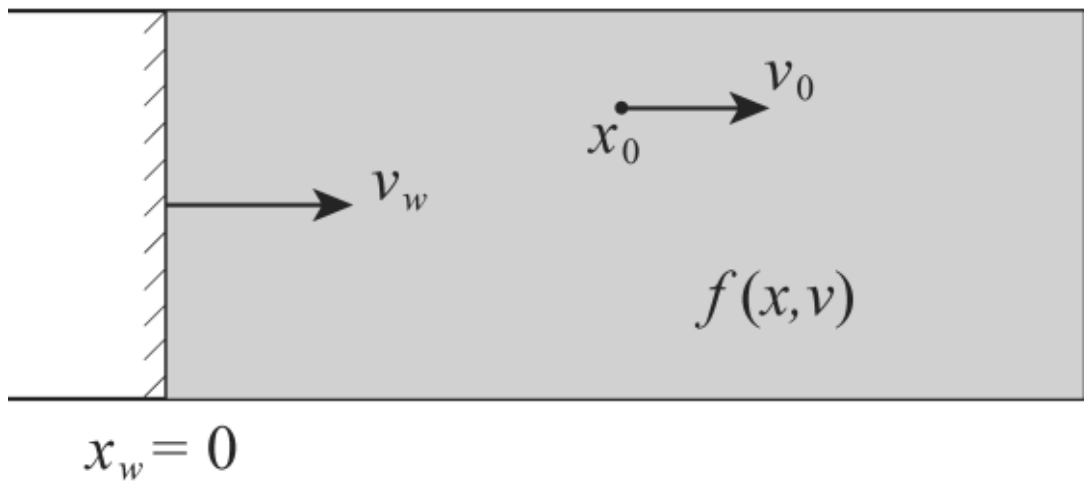

**FIG. 2.** One-dimensional moving piston.

Figure 2 shows a one-dimensional moving piston. The velocity of the piston is $v_w$ and the gas distribution function on the right side of the piston is $f\left(x,v\right)$ with $x$ and $v$ representing the particle position and velocity. At $t_0$, the piston is located at $x_w = 0$ and a particle whose velocity is $v_0$ is located at $x_0$. In the premise of $v_w > v_0$, the collision moment $t_c$ for the particle colliding with the piston is determined as below.

$$t_c = t_0 + \frac{x_0}{v_w - v_0} \tag{7}$$

If the collision occurs within the time interval $[t_0, t_1]$ ( $t_1 = t_0 + \Delta t$ ), $x_0$ must satisfy $0 \le x_0 \le \left(v_w - v_0\right)\Delta t$. Therefore, the total number of particles that will collide with the piston within the time interval, denoted $N_{\Delta t}$, is expressed in the following form.

$$N_{\Delta t} = \int_{-\infty}^{v_w} \int_0^{(v_w - v)\Delta t} f\left(x,v\right) dxdv \tag{8}$$

When the approximate algorithm illustrated in Fig. 1 is applied, this number can be re-expressed by Eq. (9) since the piston is frozen during the particle-wall collision, and an error, denoted $\Delta N_{\Delta t}$, is thus induced as shown in Eq. (10).

$$N_{\Delta t}^{*} = \int_{-\infty}^{0} \int_0^{-v\Delta t} f\left(x,v\right) dxdv \tag{9}$$

$$\Delta N_{\Delta t} = N_{\Delta t} - N_{\Delta t}^{*} = \int_{-\infty}^{v_w} \int_0^{(v_w - v)\Delta t} f\left(x,v\right) dxdv - \int_{-\infty}^{0} \int_0^{-v\Delta t} f\left(x,v\right) dxdv \tag{10}$$

As illustrated in Fig. 3, the integral domain of $\Delta N_{\Delta t}$ is the region between the lines $v = x/\Delta t$ and $v = v_w - x/\Delta t$, and it is divided into two parts by the $x$ axis, indicating that Eq. (10) can be rewritten in the following form.

$$\Delta N_{\Delta t} = \int_{0}^{v_w} \int_0^{(v_w - v)\Delta t} f\left(x,v\right) dxdv + \int_{-\infty}^{0} \int_{-v\Delta t}^{(v_w - v)\Delta t} f\left(x,v\right) dxdv \tag{11}$$

If the gas distribution is homogeneous, $f\left(x,v\right)$ is reduced to $f\left(v\right)$, and Eq. (11) can thus be simplified as below.

$$\begin{aligned} \Delta N_{\Delta t} &= \Delta t \int_0^{v_w} \left(v_w - v\right) f\left(v\right) dv + v_w \Delta t \int_{-\infty}^{0} f\left(v\right) dv \\ &= \Delta t \left[ \int_0^{v_w} \left(v_w - v\right) f\left(v\right) dv + v_w \int_{-\infty}^{0} f\left(v\right) dv \right] \end{aligned} \tag{12}$$

By letting $\Delta t \to 0$, Eq. (12) is rewritten as follows.

$$\frac{dN_{\Delta t}}{dt} = \int_0^{v_w} \left(v_w - v\right) f\left(v\right) dv + v_w \int_{-\infty}^{0} f\left(v\right) dv \tag{13}$$

Here, $dN_{\Delta t}/dt$ is the error accumulation rate of $N_{\Delta t}$. It is evident that this rate is dependent on $v_w$ but independent of $\Delta t$. Besides, $dN_{\Delta t}/dt$ increases with $v_w$. It is indicated that the error induced by the approximate algorithm cannot be reduced by shortening $\Delta t$ and its accumulation rate will become nonnegligible if $|v_w|$ is large. As a result, an accurate moving wall boundary algorithm is essentially required, especially for the case of fast-moving wall.

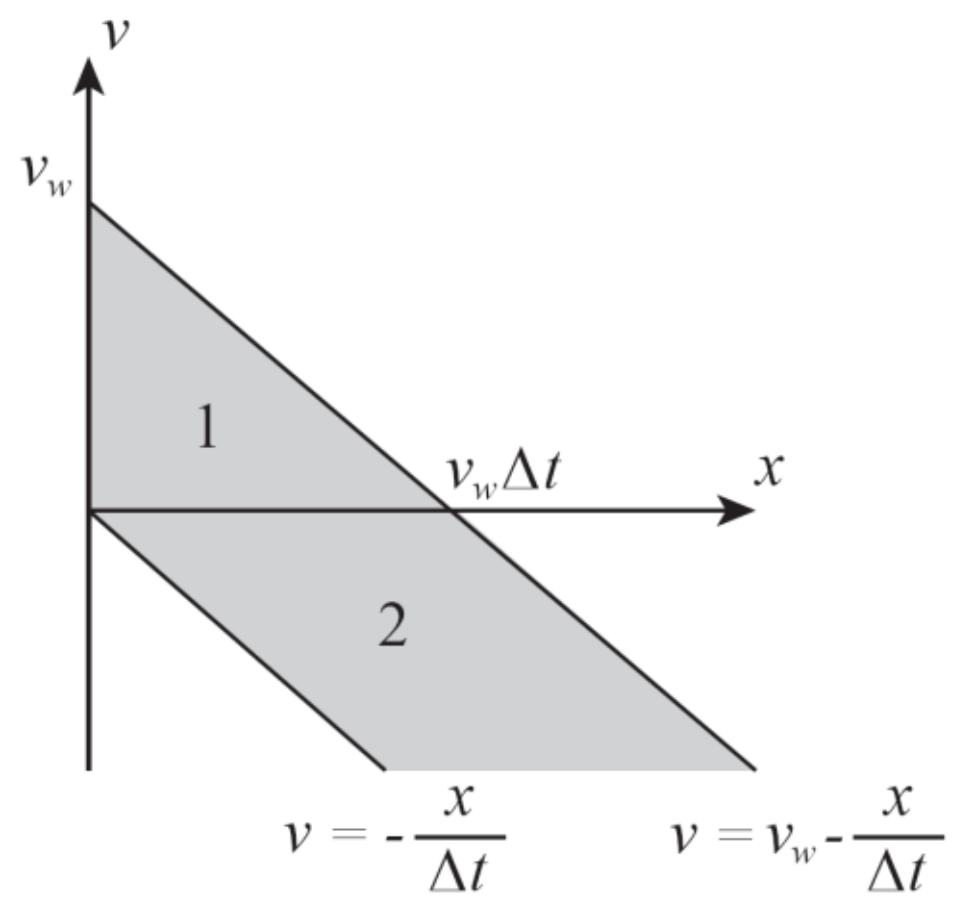

**FIG. 3.** Integral domain of $\Delta N_{\Delta t}$ in the *x*-*v* plane.

## IV. ACCURATE MOVING WALL BOUNDARY ALGORITHM

The key to avoid the error induced by the approximate algorithm is to accurately determine the position of the particle-wall collision. To this end, an accurate algorithm is proposed in the present work.

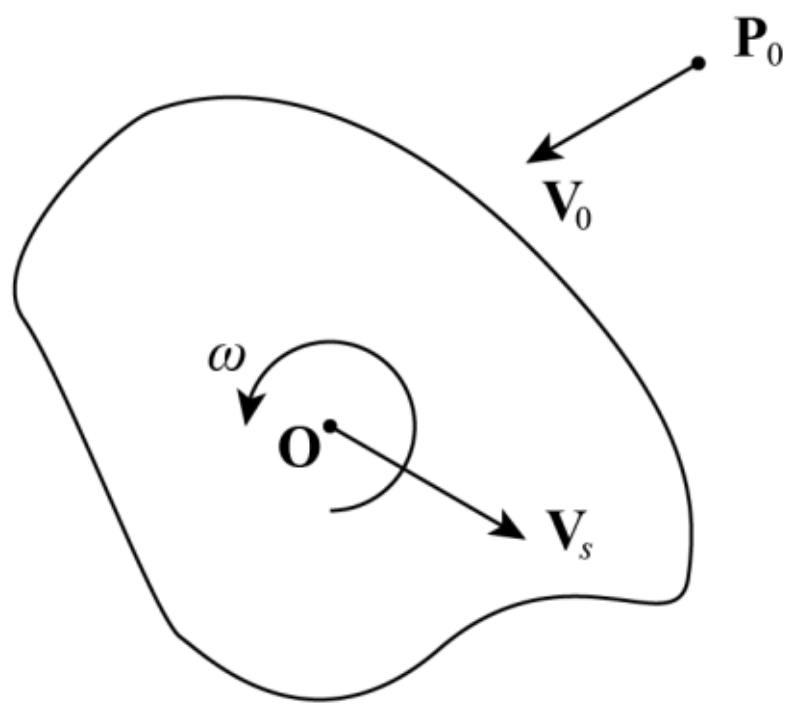

**FIG. 4.** A moving object and an incident particle.

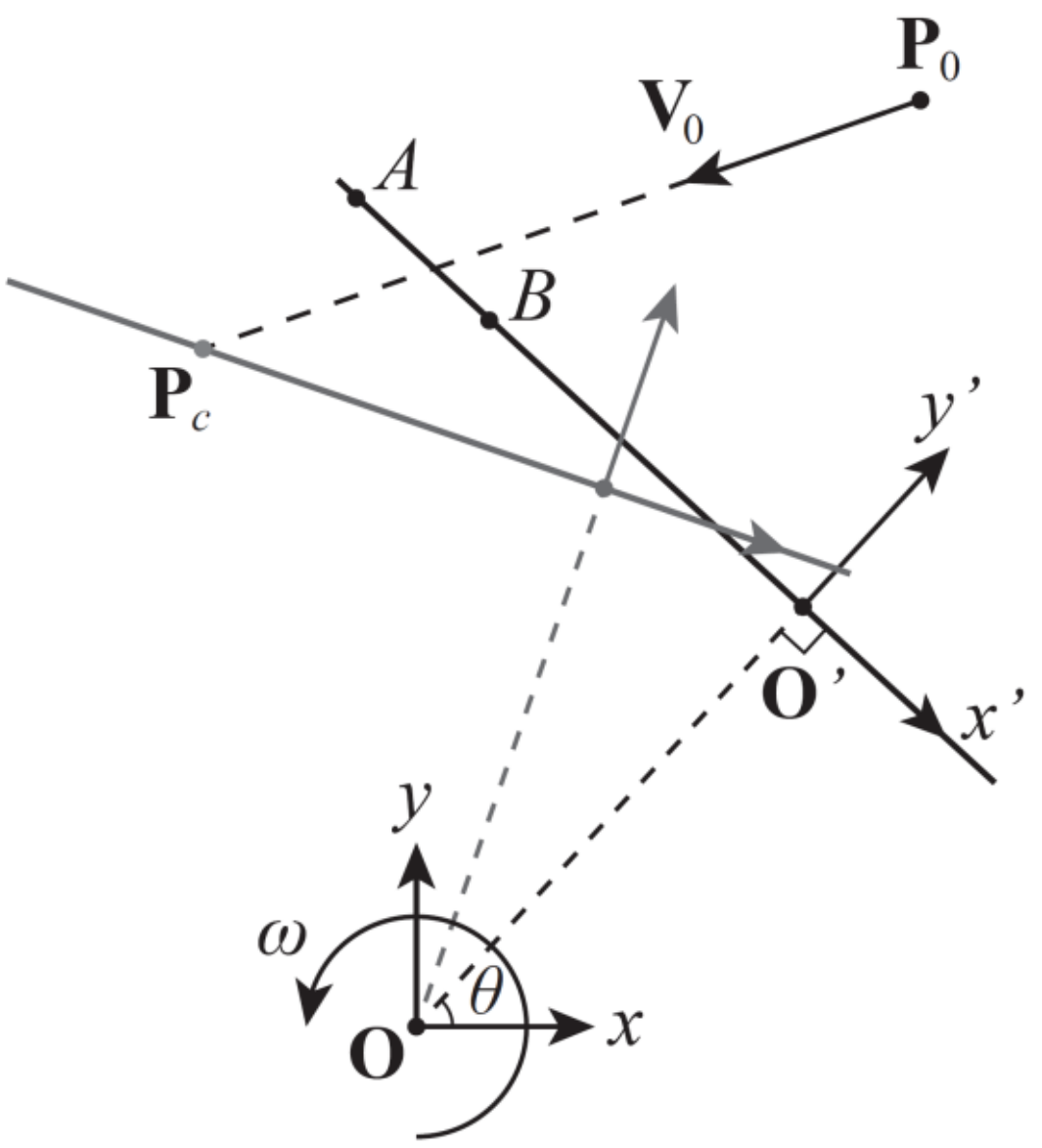

**FIG. 5.** Geometry of the particle-wall collision for the case of moving wall.

Figure 4 shows a moving object and an incident particle in two dimensions. At $t_0$, the object is translating at $\mathbf{V}_s$ and also rotating at $\omega$ around point $O$. Meanwhile, a particle located at $P_0$ is moving in a straight line at $\mathbf{V}_0$. In order to determine whether the particle trajectory passes through any wall element on the surface of the object within $[t_0,\ t_1]$ ($t_1 = t_0 + \Delta t$), a coordinate transformation is conducted by setting the coordinate origin at $O$. The transformed coordinate system is denoted $Oxy$. The particle position and velocity vectors in $Oxy$ at $t_0$, denoted respectively $\mathbf{P}_{0,Oxy}$ and $\mathbf{V}_{0,Oxy}$, are given as below.

$$\begin{aligned} \mathbf{P}_{0,Oxy} &= \mathbf{P}_0 - \mathbf{O} \\ \mathbf{V}_{0,Oxy} &= \mathbf{V}_0 - \mathbf{V}_s \end{aligned} \tag{14}$$

Here, $\mathbf{O}$ is the position vector of $O$ in the inertial coordinate system at $t_0$. Before particle colliding with the object, the particle trajectory in $Oxy$ can be expressed as a function of time as shown in Eq. (15).

$$\mathbf{P}_{Oxy}(t) = \mathbf{P}_{0,Oxy} + (t - t_0)\mathbf{V}_{0,Oxy} \tag{15}$$

Here, $\mathbf{P}_{Oxy}$ is the particle position vector in $Oxy$ at $t$. As illustrated in Fig. 5, the line segment $AB$ is a wall element on the surface of the object and its motion in $Oxy$ is the rotation around $O$. There is always a point on $AB$ or on its extension that is closest to $O$. This point is denoted $O'$ and the line segment $OO'$ is perpendicular to $AB$. The angle between the $x$ axis and $OO'$ is $\theta$ and is a function of time as given below.

$$\theta(t)=\theta_0+\omega(t-t_0) \tag{16}$$

Here, $\theta_0$ is the angle at $t_0$. Therefore, the coordinates of $O'$, which are also functions of time, can be expressed in terms of $\theta$ in the following forms.

$$\begin{aligned} x_{O'}(t) &= r\cos\left[\theta(t)\right] \\ y_{O'}(t) &= r\sin\left[\theta(t)\right] \end{aligned} \tag{17}$$

Here, $r$ is the length of $OO'$. By setting $O'$ as the origin as well as setting the $x'$ and $y'$ axes in the directions of $\overrightarrow{AB}$ and $\overrightarrow{OO'}$, a new coordinate system, denoted $O'x'y'$, is constructed as shown in Fig. 5. In $O'x'y'$, Eq. (15) is re-expressed as follows.

$$\mathbf{P}_{O'x'y'}(t)=\begin{bmatrix} \sin\left[\theta(t)\right] & -\cos\left[\theta(t)\right] \\ \cos\left[\theta(t)\right] & \sin\left[\theta(t)\right] \end{bmatrix}\left[\mathbf{P}_{Oxy}(t)-\mathbf{O}'\right] \tag{18}$$

Here, $\mathbf{P}_{O'x'y'}$ is the particle position vector in $O'x'y'$ at $t$ and $\mathbf{O}'$ is the position vector of $O'$ in $Oxy$ at $t$. Then, the particle trajectory in $O'x'y'$ can be determined by substituting Eqs. (14), (15) and (17) into Eq. (18). The components of the particle trajectory are given below.

$$\begin{aligned} x_{\mathbf{P}_{O'x'y'}}(t) &= \left\{x_{\mathbf{P}_{Oxy}}(t)-r\sin\left[\theta(t)\right]\right\}\sin\left[\theta(t)\right]-\left\{y_{\mathbf{P}_{Oxy}}(t)-r\cos\left[\theta(t)\right]\right\}\cos\left[\theta(t)\right] \\ y_{\mathbf{P}_{O'x'y'}}(t) &= \left\{x_{\mathbf{P}_{Oxy}}(t)-r\sin\left[\theta(t)\right]\right\}\cos\left[\theta(t)\right]+\left\{y_{\mathbf{P}_{Oxy}}(t)-r\cos\left[\theta(t)\right]\right\}\sin\left[\theta(t)\right] \end{aligned} \tag{19}$$

If the particle trajectory has a geometric intersection with $AB$, a root of the following equation within $[t_0,\ t_1]$ must exist.

$$y_{\mathbf{P}_{O'x'y'}}(t)=0 \tag{20}$$

Besides, the point, corresponding to this root, must be on $AB$.

Eq. (20) can be solved by the Newton iteration method and the solution is treated as the collision moment $t_c$. As long as the solution of Eq. (20) is obtained, the particle position at $t_1$ can be accurately determined by Eq. (5).

Although the above derivation is based on the two-dimensional case, it can be extended to three dimensions. Eq. (21) shows the expression of the vector form of the particle trajectory in three dimensions.

$$\mathbf{P}_{O'x'y'z'}(t)=\mathbf{M}\left[\mathbf{P}_{Oxyz}(t)-\mathbf{O}'\right] \tag{21}$$

Here, $\mathbf{M}$, similar to that in Eq. (18), is a linear transformation matrix, indicating the rotation around the axis of rotation. The form of $\mathbf{M}$ is given as follows.

$$\mathbf{M}=\begin{bmatrix} \cos\theta+(1-\cos\theta)A_x^2 & (1-\cos\theta)A_xA_y-\sin\theta A_z & (1-\cos\theta)A_xA_z\sin\theta+A_y \\ (1-\cos\theta)A_xA_y+\sin\theta A_z & \cos\theta+(1-\cos\theta)A_y^2 & (1-\cos\theta)A_yA_z\sin\theta-A_x \\ (1-\cos\theta)A_xA_z-\sin\theta A_y & (1-\cos\theta)A_yA_z+\sin\theta A_x & \cos\theta+(1-\cos\theta)A_z^2 \end{bmatrix} \quad (22)$$

Here, $\theta$ is the rotation angle around the axis of rotation and will be known as long as the manner of rotation is known or can be predicted; $A_x$, $A_y$ and $A_z$ are the unit directional vector components of the axis of rotation. If the axis of rotation is fixed, $A_x$, $A_y$ and $A_z$ can be straightforwardly obtained. If it is unfixed, $A_x$, $A_y$ and $A_z$ will vary with time. In such case, they can be reset in each time step as long as the manner of rotation is known or can be predicted. For this reason, the algorithm can handle the case of unfixed axis of rotation.

## V. RESULTS AND DISCUSSIONS

In this section, the accurate algorithm is validated in four rarefied flow cases. The first is the one-dimensional normal shock wave formation, the second is the flow inside a rotating square box, and the last two are unsteady flows around a periodically rotating ellipse and the three-dimensional Apollo command Module.

### A. One-dimensional normal shock wave formation

As shown in Fig. 2, the piston moves and compresses the gas (argon). If the velocity of the piston exceeds the local sonic speed, a normal shock wave will form in front of the piston. This is a typical unsteady problem involving a moving wall and was simulated by Bird[16] using the piston boundary algorithm within the DSMC framework.

The theoretical solutions of the post-shock state can be obtained from the Rankin-Hugoniot relations as follows.

$$\begin{aligned} \frac{\rho_2}{\rho_1}&=\frac{v_1}{v_2}=\frac{(\gamma+1)M_a^2}{2+(\gamma-1)M_a^2} \\ \frac{p_2}{p_1}&=1+\frac{2\gamma}{\gamma+1}\left(M_a^2-1\right) \\ \frac{T_2}{T_1}&=\left[1+\frac{2\gamma}{\gamma+1}\left(M_a^2-1\right)\right]\frac{2+(\gamma-1)M_a^2}{(\gamma+1)M_a^2} \end{aligned} \quad (23)$$

Here, $\rho$, $v$, $p$ and $T$ are respectively the density, velocity, pressure and temperature, subscripts 1 and 2 indicate respectively the pre- and post-shock states, $\gamma$ is the specific heat ratio of the gas, and $M_a$ is the Mach number of the incoming flow. According to Eq. (23), the relation between the piston velocity $v_w$ and the normal shock wave velocity $v_s$ is derived as below.

$$\frac{v_w}{v_s} = \frac{v_2 - v_1}{-v_1} = 1 - \frac{2 + (\gamma - 1) M_s^2}{(\gamma + 1) M_s^2} \tag{24}$$

Here, $M_s$ is the Mach number corresponding to $v_s$. In the present work, $v_w$, $T_1$ and $\rho_1$ are set to 1,818 m/s, 273 K and $3.34\times10^{-7}$ kg/m$^3$, respectively.

The calculation domain is 100 m long. It is discretized into 2,000 grids with each grid 0.05 m in length. The time step is $5.0\times10^{-6}$ s. The scaling factor that indicates the number of molecules represented by a particle is $1.0\times10^{13}$. The number of particles is $5.0\times10^{7}$. In order to smooth fluctuations, the ensemble average is conducted using the results from 20 computations.

Figure 6 displays the temporal variations of the normalized density and temperature profiles obtained using the accurate algorithm. The black solid line indicates the trajectory of the piston and the black dashed line indicates the theoretical trajectory of the normal shock wave determined by Eq. (24). It is shown that the accurate algorithm is able to capture the shock wave because the black dashed line is located in the region where the density and temperature vary dramatically. It is also shown that the density and temperature near the piston are slightly deviated from the post-shock state determined by Eq. (23). As explained by Bird,[16] it is caused by an entropy layer attached to the piston, which is generated by gas passing through the incomplete normal shock wave when the piston starts to move from the stationary state.

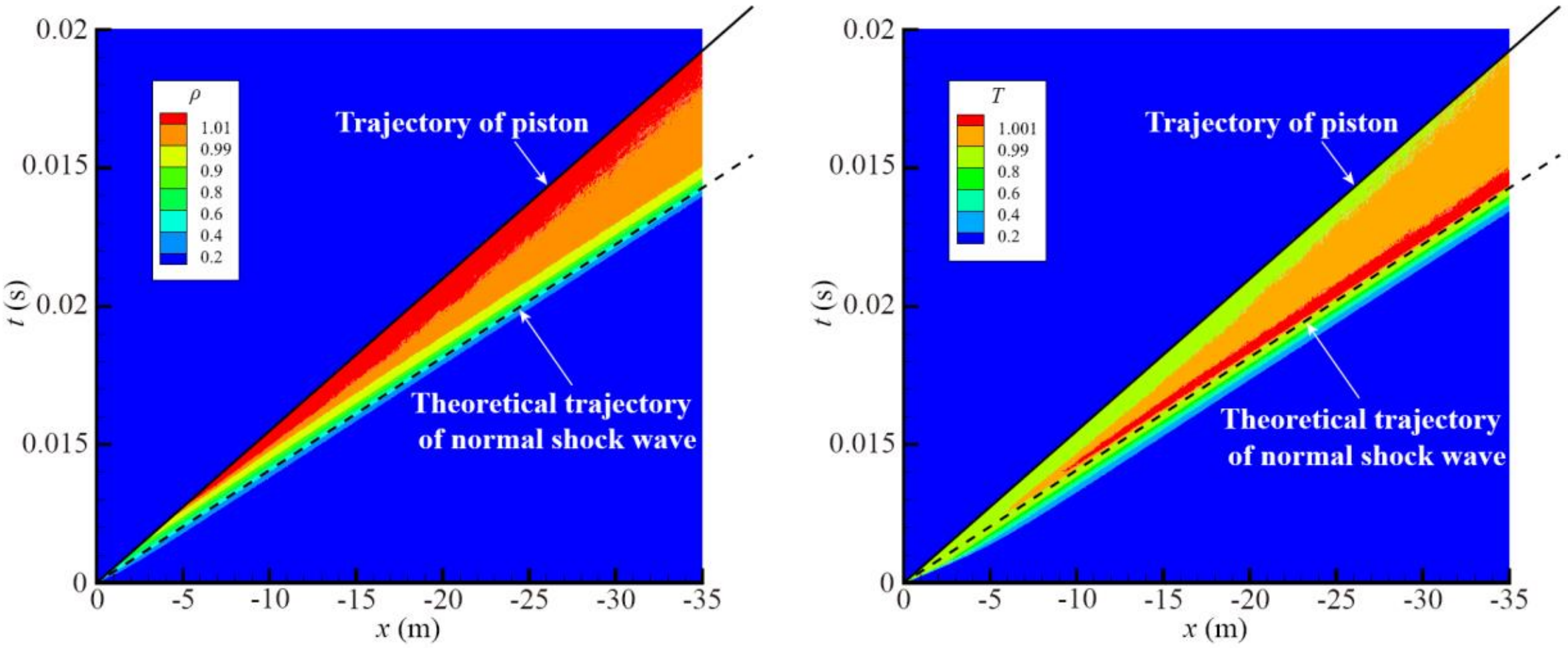

**FIG. 6.** Temporal variations of the normalized density (left) and temperature (right) profiles obtained using the accurate algorithm for the one-dimensional normal shock wave formation.

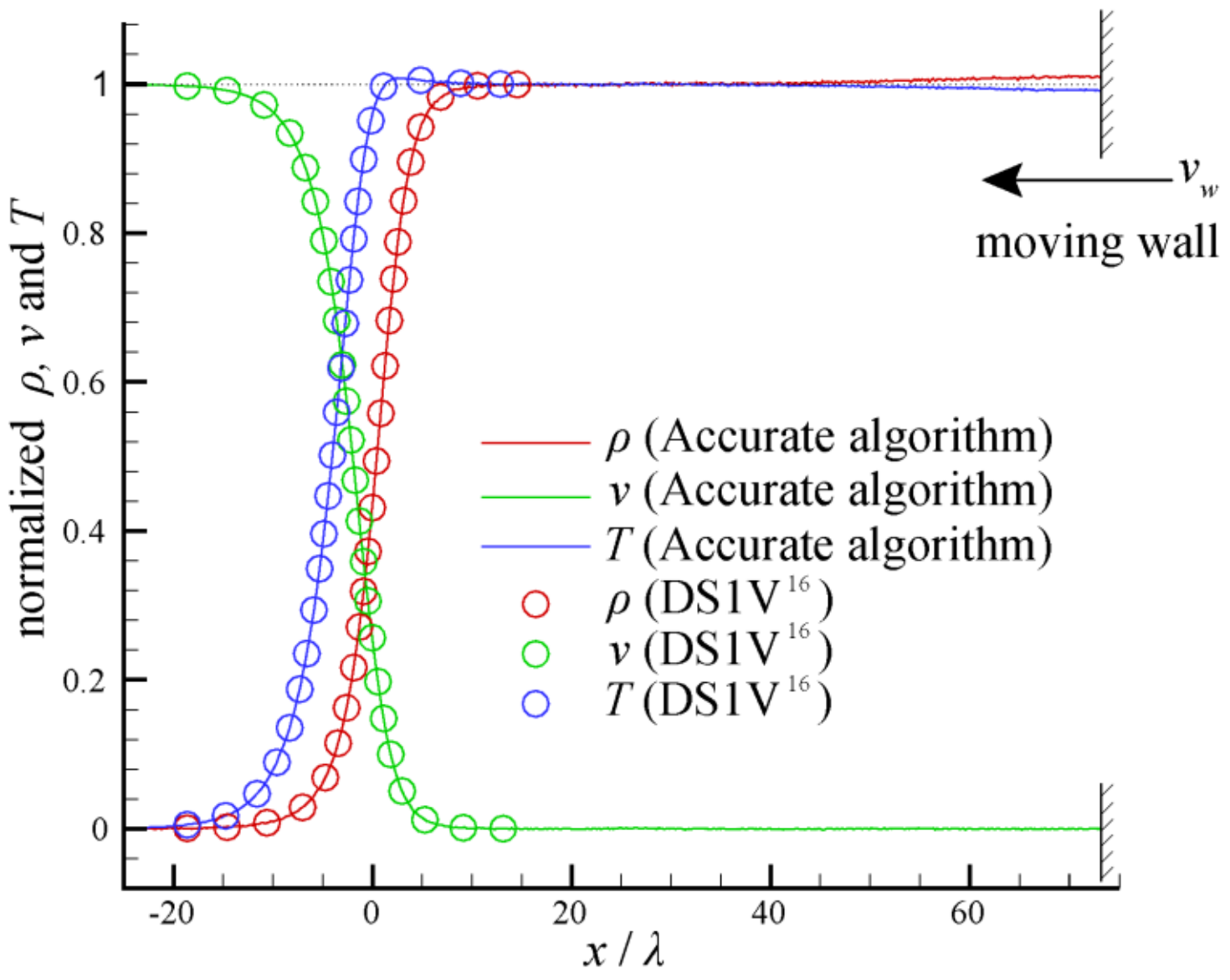

**FIG. 7.** Normalized density, velocity and temperature profiles at 0.03 s for the one-dimensional normal shock wave formation.

Given in Fig. 7 are the normalized density, velocity and temperature profiles at 0.03 s. The $x$ coordinate in this figure is nondimensionalized by the upstream mean free path $\lambda$ that is defined in the following form.

$$\lambda = \frac{1}{\sqrt{2} n \pi d^2} \tag{25}$$

Here, $n$ and $d$ are respectively the number density and diameter of molecules. It is seen that the profiles are in good agreement with that obtained using the DS1V code,[16] thus validating the accurate algorithm on a quantitative basis. It is also seen that there are slight deviations of the density and temperature at the near-wall region. This is caused by the aforementioned effect of the entropy layer. Considering that the density is only 1.05 % greater and the temperature is only 0.96 % lower than the post-shock state, the entropy layer will have little effect on the aerodynamic properties upon the surface.

The approximate algorithm is also adopted in the simulation. It is exhibited in Fig. 8 that the normalized density predicted by this algorithm is inaccurate and severely discrepant from the physics. It is because the approximate algorithm decouples the motions of wall and particles by freezing wall during the particle-wall collision, it cannot identify each collision and tends to induce particle to penetrate wall and leak out before the collision happens. The piston moves at a very high speed (1,818 m/s), which means that using the approximate algorithm will induce most of particles to penetrate the piston, leading to a severe gas leakage that gives rise to the nonphysical prediction. In contrast, the accurate algorithm takes the effect of wall's motion into consideration during the collision and thus can accurately determine the collision moment and position. Therefore, it can avoid the gas leakage and ensure accuracy as shown in Fig. 7.

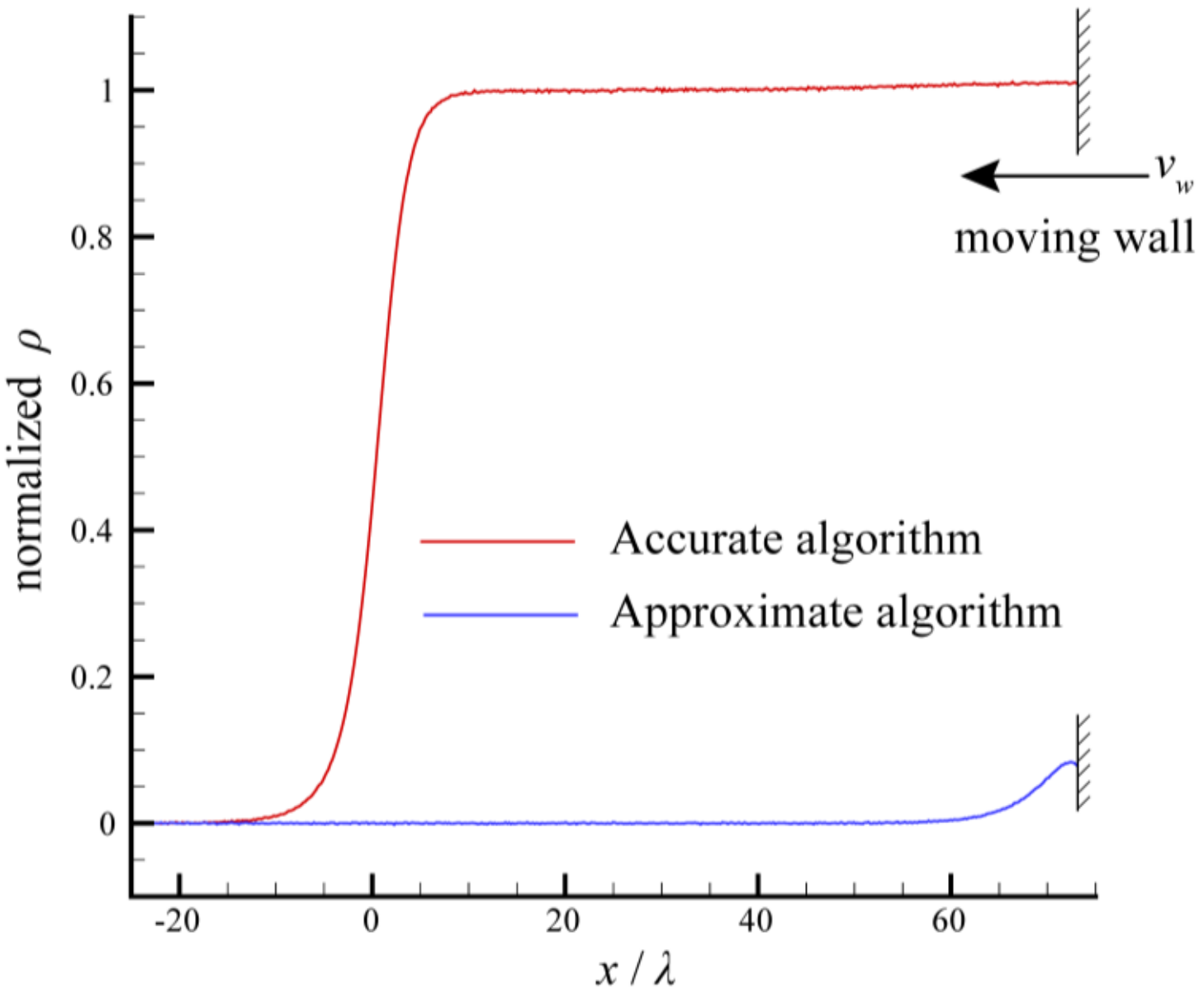

**FIG. 8.** Normalized density profile at 0.03 s obtained using the accurate and approximate algorithms for the one-dimensional normal shock wave formation.

It is noted that the one-dimensional piston boundary algorithm[16] is the reduced case of the accurate algorithm and is able to accurately describe the particle-wall collision in cases involving only translational motion. However, it is unable to deal with cases involving rotational motion, while the accurate algorithm, in contrast, is able to do that as shown in the following sections.

### B. Rotating square box

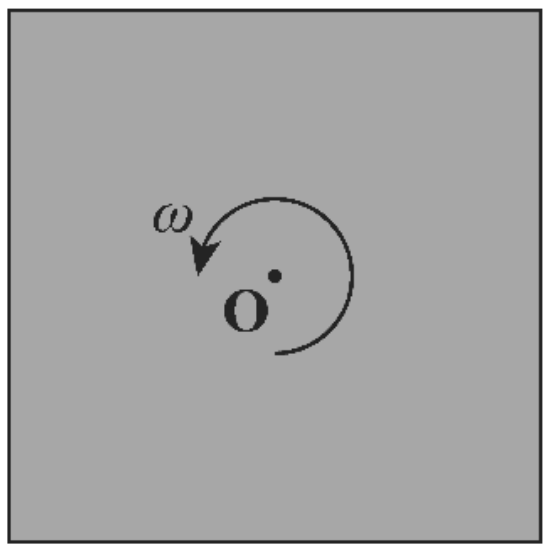

**FIG. 9.** Rotating square box.

Figure 9 shows a square box rotating around its center at $50\pi$ rad/s. The side length, the wall temperature and the wall accommodation coefficient of the box are 5 m, 273 K and 1, respectively. At the beginning, the gas (argon) is homogeneously distributed inside the box and its density and temperature are respectively $3.34\times10^{-7}$ kg/m$^3$ and 273 K. After a sufficient time, the gas rotates in equilibrium. Maxwell[38] proposed a relation, as shown in Eq. (26), to describe the gas distribution at the equilibrium state for this rotating case.

$$f\left(x,y,u,v,w\right)=\rho_O\left(\frac{m}{2\pi k_B T_O}\right)^{\frac{3}{2}}\exp\left\{-\frac{m}{2k_B T_O}\left[u^2+v^2+w^2+2\omega\left(uy-vx\right)\right]\right\} \tag{26}$$

Here, $\omega$ is the angular velocity of the box; $\rho_O$ and $T_O$ are respectively the density and temperature at point $O$ that is at the center of the box; $u$, $v$ and $w$ are respectively the *x*, *y* and *z*-direction velocities of molecules. By integrating Eq. (26), the distributions of the density $\rho\left(x,y\right)$ and temperature $T\left(x,y\right)$ as well as velocities $U\left(x,y\right)$ and $V\left(x,y\right)$ at the equilibrium state are determined as follows.

$$\begin{aligned}
&\rho\left(x,y\right)=\int_{-\infty}^{\infty}\int_{-\infty}^{\infty}\int_{-\infty}^{\infty}f\left(x,y,u,v,w\right)dudvdw=\rho_O\exp\left[\frac{m\omega^2}{2k_B T_O}\left(x^2+y^2\right)\right]\\
&T\left(x,y\right)=\frac{1}{\rho}\int_{-\infty}^{\infty}\int_{-\infty}^{\infty}\int_{-\infty}^{\infty}\left[\left(u-U\right)^2+\left(v-V\right)^2+w^2\right]\frac{m}{3k_B}f\left(x,y,u,v,w\right)dudvdw=T_O\\
&U\left(x,y\right)=\frac{1}{\rho}\int_{-\infty}^{\infty}\int_{-\infty}^{\infty}\int_{-\infty}^{\infty}uf\left(x,y,u,v,w\right)dudvdw=-\omega y\\
&V\left(x,y\right)=\frac{1}{\rho}\int_{-\infty}^{\infty}\int_{-\infty}^{\infty}\int_{-\infty}^{\infty}vf\left(x,y,u,v,w\right)dudvdw=\omega x
\end{aligned} \tag{27}$$

It is indicated by Eq. (27) that the gas is stationary relative to the box and has a spatially uniform temperature distribution at the equilibrium state. Due to the conservation of mass, the spatial integral of $\rho\left(x,y\right)$ satisfies the following equation.

$$\int_{-l/2}^{l/2}\int_{-l/2}^{l/2}\rho_O\exp\left[\frac{m\omega^2}{2k_B T_O}\left(x^2+y^2\right)\right]dxdy=\rho_0 l^2 \tag{28}$$

Here, $l$ is the side length of the box and $\rho_0$ is the uniform density of the gas at the beginning. By solving Eq. (28), $\rho_O$ is obtained as given below.

$$\rho_O=\frac{m\rho_0 l^2\omega^2}{4\pi k_B T\left[\operatorname{erfi}\left(\sqrt{\frac{ml^2\omega^2}{8k_B T}}\right)\right]^2}=1.136\times10^{-7}\ \mathrm{kg/m^3} \tag{29}$$

Here, erfi indicates the imaginary error function.

The box is placed in the center of the calculation domain that is 10 m×10 m in size. The calculation domain is discretized into 200×200 quadrate grids. Each grid is 0.05 m in size. The time step is $5.0\times10^{-6}$ s. The scaling factor and the number of particles are $3.0\times10^{14}$ and $1.68\times10^{6}$, respectively. The computational results are temporally averaged over 200 cycles to smooth fluctuations.

Figure 10 displays the equilibrium density distribution obtained using the accurate algorithm (on the left side of the black dashed line) compared with that of the Maxwellian theoretical solution determined by Eq. (27) (on the right side of the black dashed line). It is clearly seen that the simulated result agrees well with the theoretical solution.

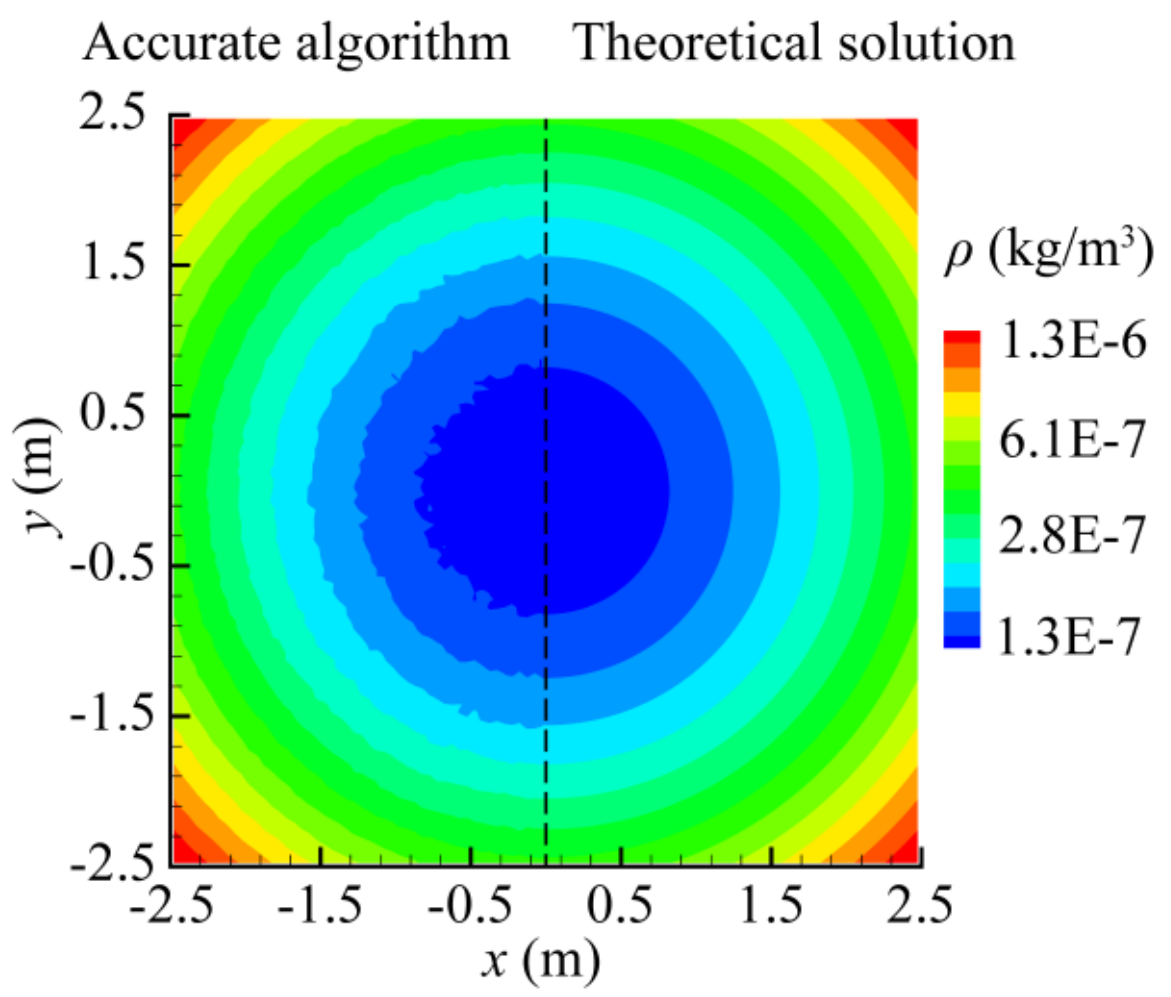

**FIG. 10.** Equilibrium density distribution inside the rotating square box.

Figure 11 summarizes and compares the profiles of the equilibrium density, temperature and $y$-direction velocity at $y$ = 0 between the results produced by the accurate algorithm and the Maxwellian theoretical solutions. It can be seen that the accurate algorithm is able to produce results that are perfectly consistent with the theoretical solutions. In addition, the temperature of the present simulation at $y$ = 0 is approximately 273 K, thus conforming the fact that the temperature distribution at the equilibrium state is uniform. Besides, it is noted that there is no velocity slip on the box wall, this is due to the fact that the gas is relatively stationary to the box at the equilibrium state as shown in Eq. (27).

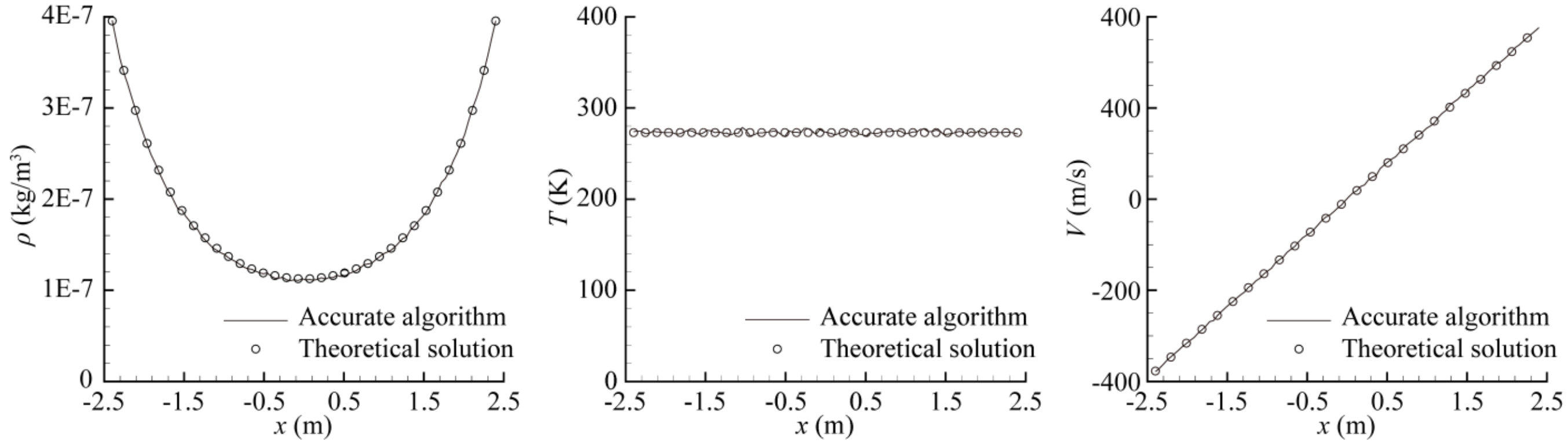

**FIG. 11.** Profiles of the equilibrium density (left), temperature (middle) and $y$-direction velocity at $y$ = 0 (right) inside the rotating square box.

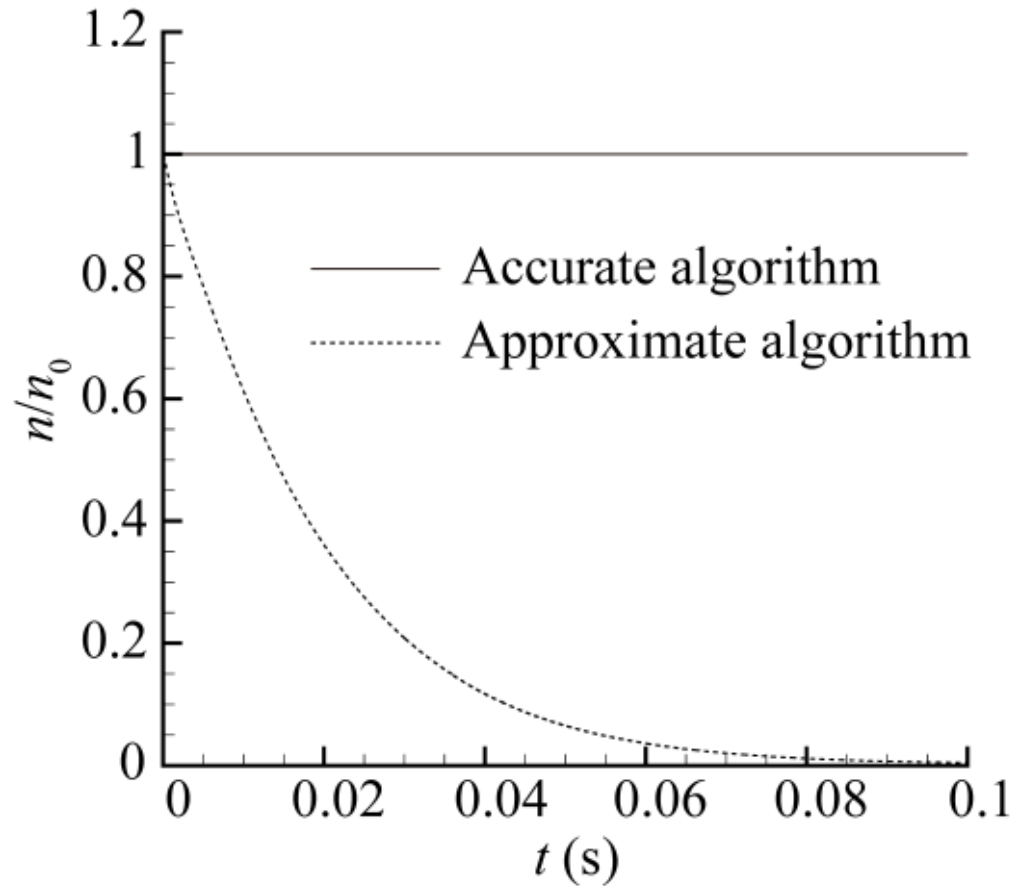

**FIG. 12.** Temporal variation of the particle number inside the rotating square box.

Because the rotating square box is a closed system and the particle number inside keeps unchanged, the comparison between the accurate and approximate algorithms in the prediction of the particle number, as shown in Fig. 12, will illustrate the quantitative discrepancy between the two algorithms. In this figure, $n$ and $n_0$ denote the instantaneous and initial particle numbers, respectively. It is evident that the accurate algorithm ensures particle conservation, while the approximate one induces gas leakage, in that the particles almost vanish from the box after 0.1 s. For this reason, the approximate algorithm produces inaccurate result (the equilibrium density is 0) that is severely discrepant from the physics.

### C. Periodically rotating ellipse

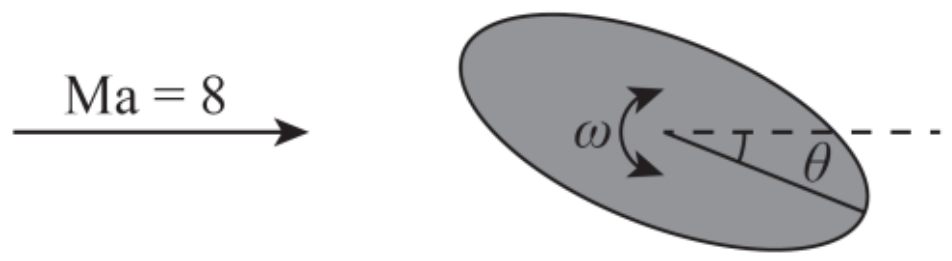

**FIG. 13.** Periodically rotating ellipse.

In order to validate the accurate algorithm in complex problems, a periodically rotating ellipse as shown in Fig. 13 is employed as a case in this work. The ellipse rotates around its center in a Mach 8 flow (argon) and its rotation angle $\theta$ is a sinusoidal function of time as given below.

$$\theta=\theta_{\max}\sin\left(2\pi f_p t\right) \tag{30}$$

Here, $f_p$ is the rotation frequency; $\theta_{\max}$ is the rotation amplitude and set to 45°. The lengths of the long and shot axes of the ellipse are 6 m and 3 m, respectively. The wall temperature is 273 K and the wall accommodation coefficient is 0.8. The density and temperature of the incoming flow are $3.34\times10^{-7}$ kg/m$^3$ and 273 K, respectively.

The calculation domain is comprised of 400,000 quadrate grids with each grid 0.05 m in size. The time step is also $5.0\times10^{-6}$ s. The scaling factor and the number of particles are respectively $3.0\times10^{14}$ and $2.35\times10^{6}$. The computational results are temporally averaged over 100 cycles.

Figures 14 and 15 display the simulated pressure distributions at $f_p$ = 31.25 Hz and 125 Hz, respectively. It is noted that the approximate algorithm tends to produce lower pressure in front of the ellipse compared to the accurate one. The maximum pressures at different moments for the two algorithms are shown in Table I, where $t_p$ is the rotation time period. For the case of 31.25 Hz, it is seen that the maximum pressure at $t$ = 0 and $0.125t_p$ predicted by the approximate algorithm are respectively 5.5% and 1.2% lower than that by the accurate one. This is an inevitable result since the approximate algorithm with the frozen-wall assumption is unable to accurately determine the position of the particle-wall collision, and thus induces particles to penetrate the wall, leading to a pressure decrease. For the case of 125 Hz, the pressure decrease becomes more server. It is consistent with the indication of Eq. (13) in section III, in that the error induced by the approximate algorithm grows with wall velocity.

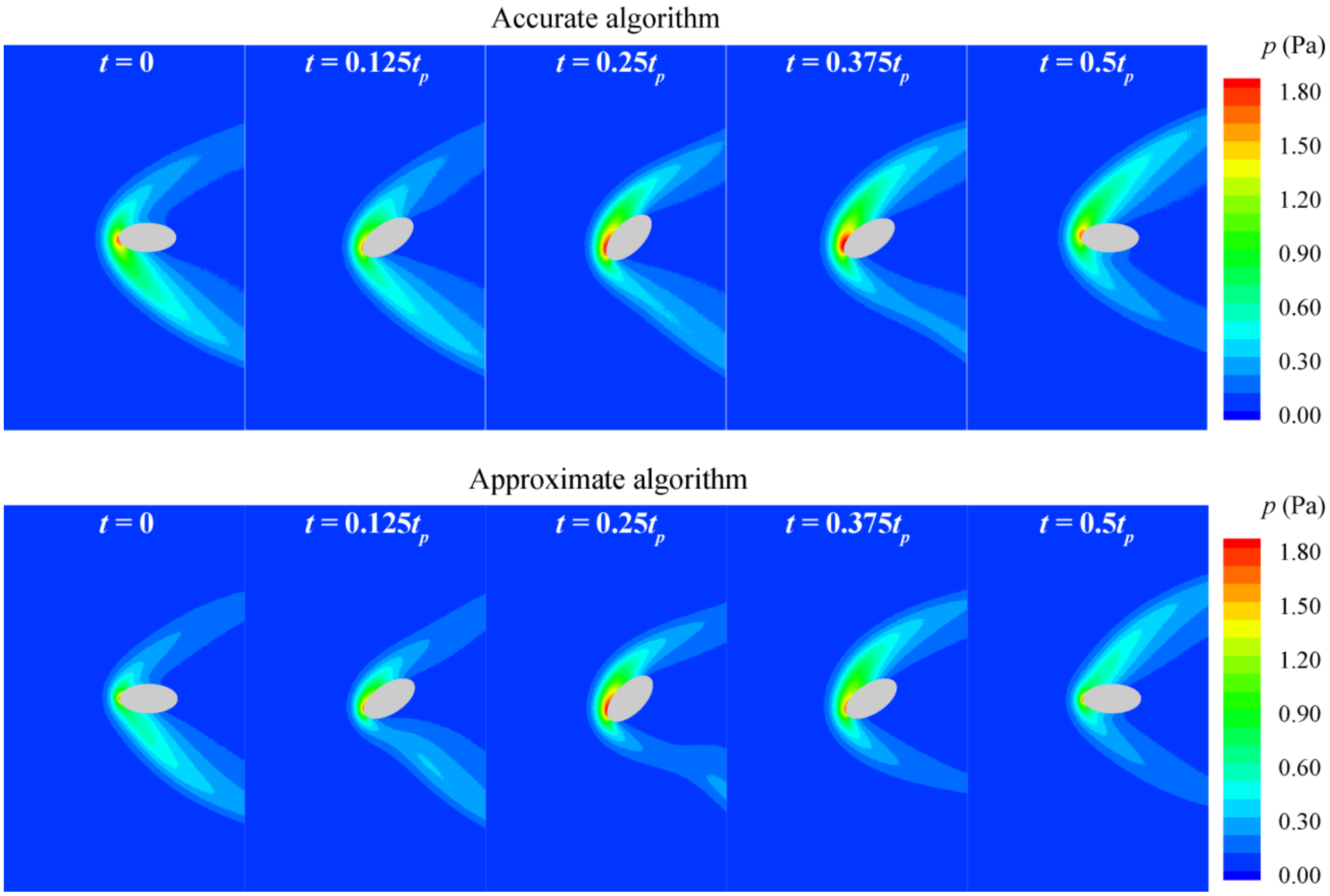

**FIG. 14.** Pressure distributions obtained using the accurate (top) and approximate (bottom) algorithms for the periodically rotating ellipse ($f_p$ = 31.25 Hz).

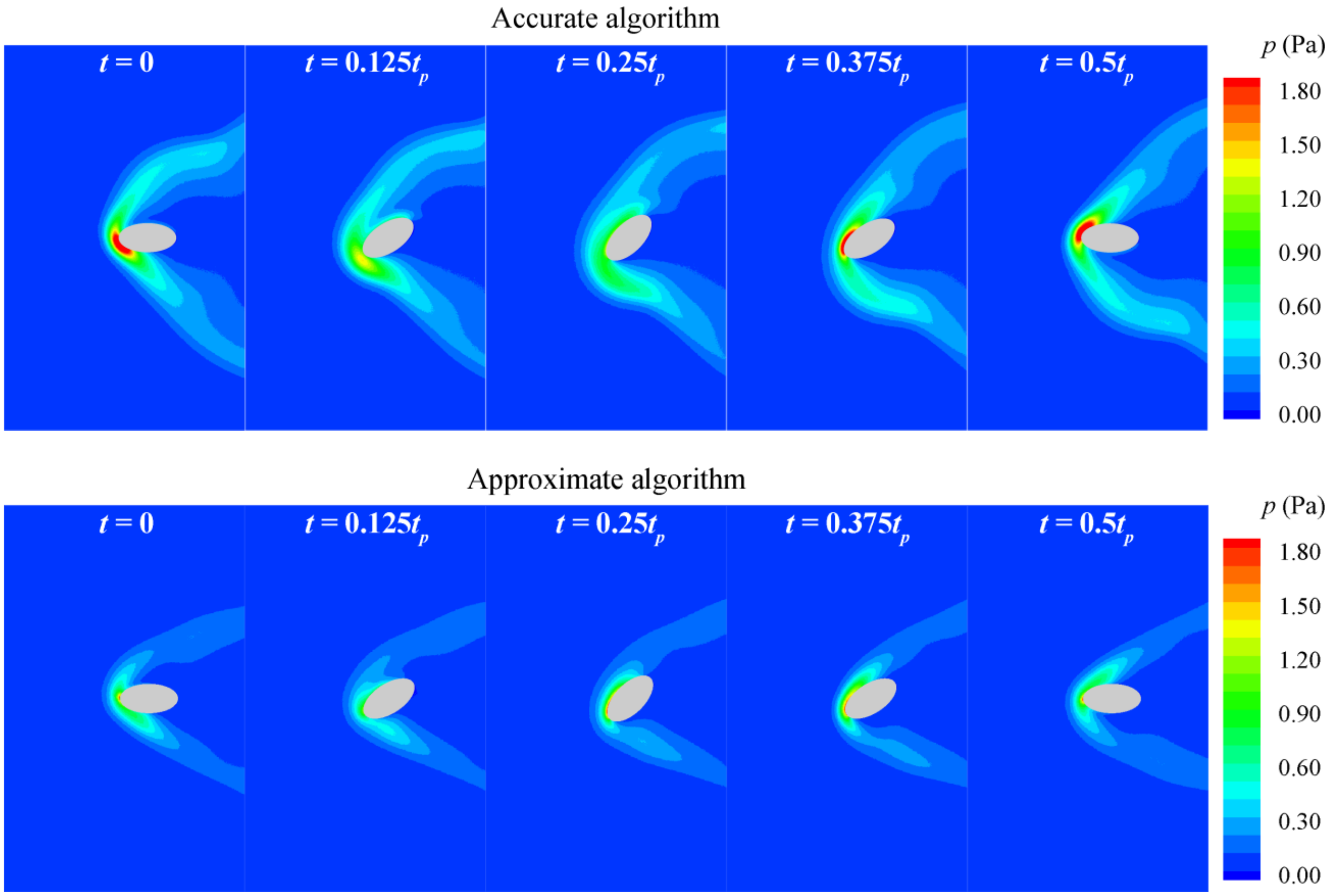

**FIG. 15.** Pressure distributions obtained using the accurate (top) and approximate (bottom) algorithms for the periodically rotating ellipse ($f_p$ = 125 Hz).

**TABLE I.** Maximum pressure obtained using the accurate and approximate algorithms for the periodically rotating ellipse.

| $f_p$ (Hz) | $t$ (s) | Maximum pressure (Pa) | | Deviation (%) |
|---|---|---|---|---|
| | | Accurate algorithm | Approximate algorithm | |
| 31.25 | 0 | 1.83 | 1.73 | 5.5 |
| | $0.125t_p$ | 1.67 | 1.65 | 1.2 |
| 125 | 0 | 2.84 | 1.84 | 35.2 |
| | $0.125t_p$ | 1.43 | 1.15 | 19.6 |

The gas leakage induced by the approximate algorithm is revealed by the particle distributions in Figs. 16 and 17. Only 1/1,000 of the particles are shown in the figures. It is evident that the approximate algorithm induces particles to penetrate the rotating wall and enter the ellipse. Furthermore, more particles enter the ellipse at the higher rotation frequency (125

Hz), indicating a more severe gas leakage. In contrast, the accurate algorithm ensures that the particle-wall collision is accurately predicted and thus avoids gas leakage.

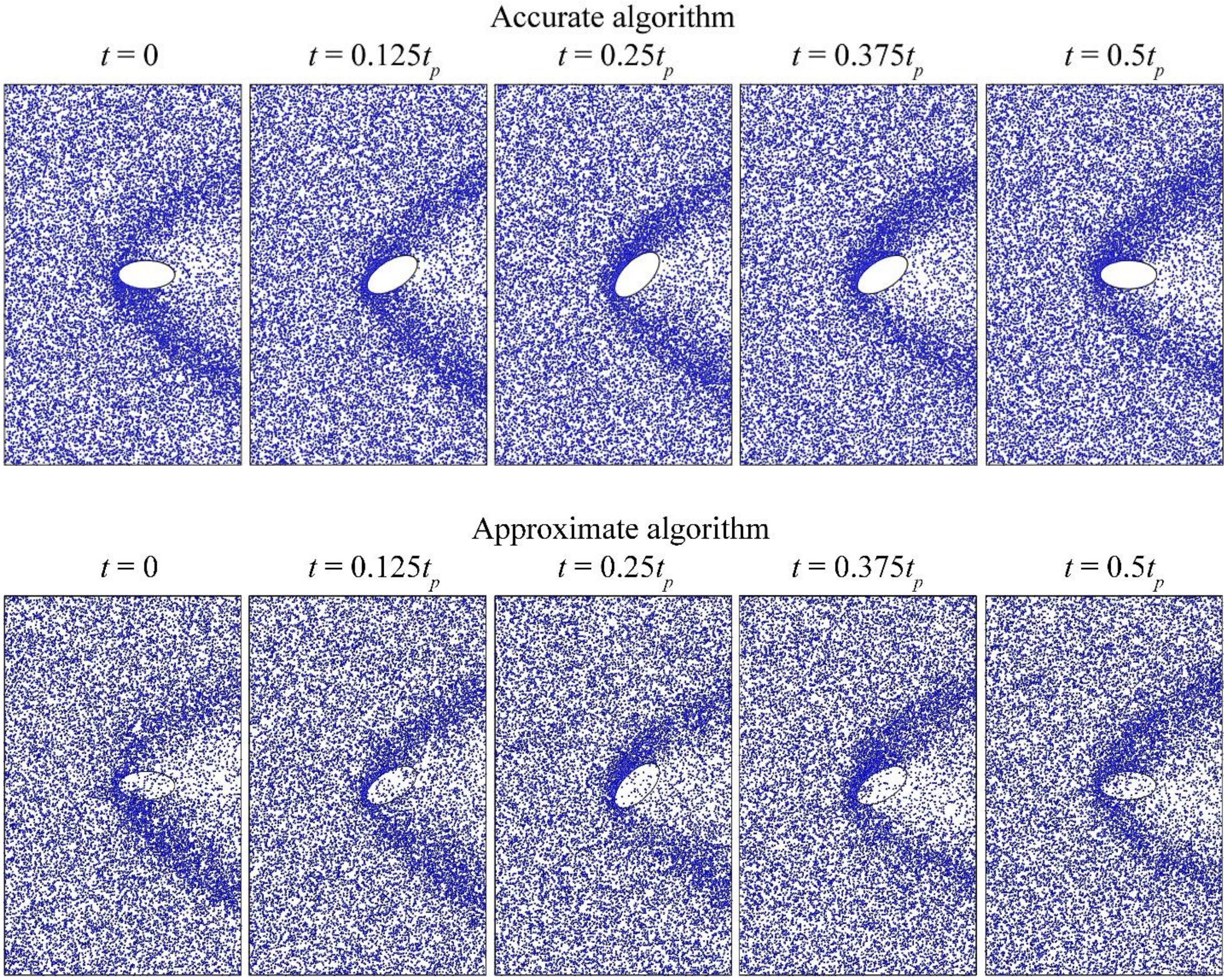

**FIG. 16.** Particle distributions obtained using the accurate (top) and approximate (bottom) algorithms for the periodically rotating ellipse ($f_p$ = 31.25 Hz).

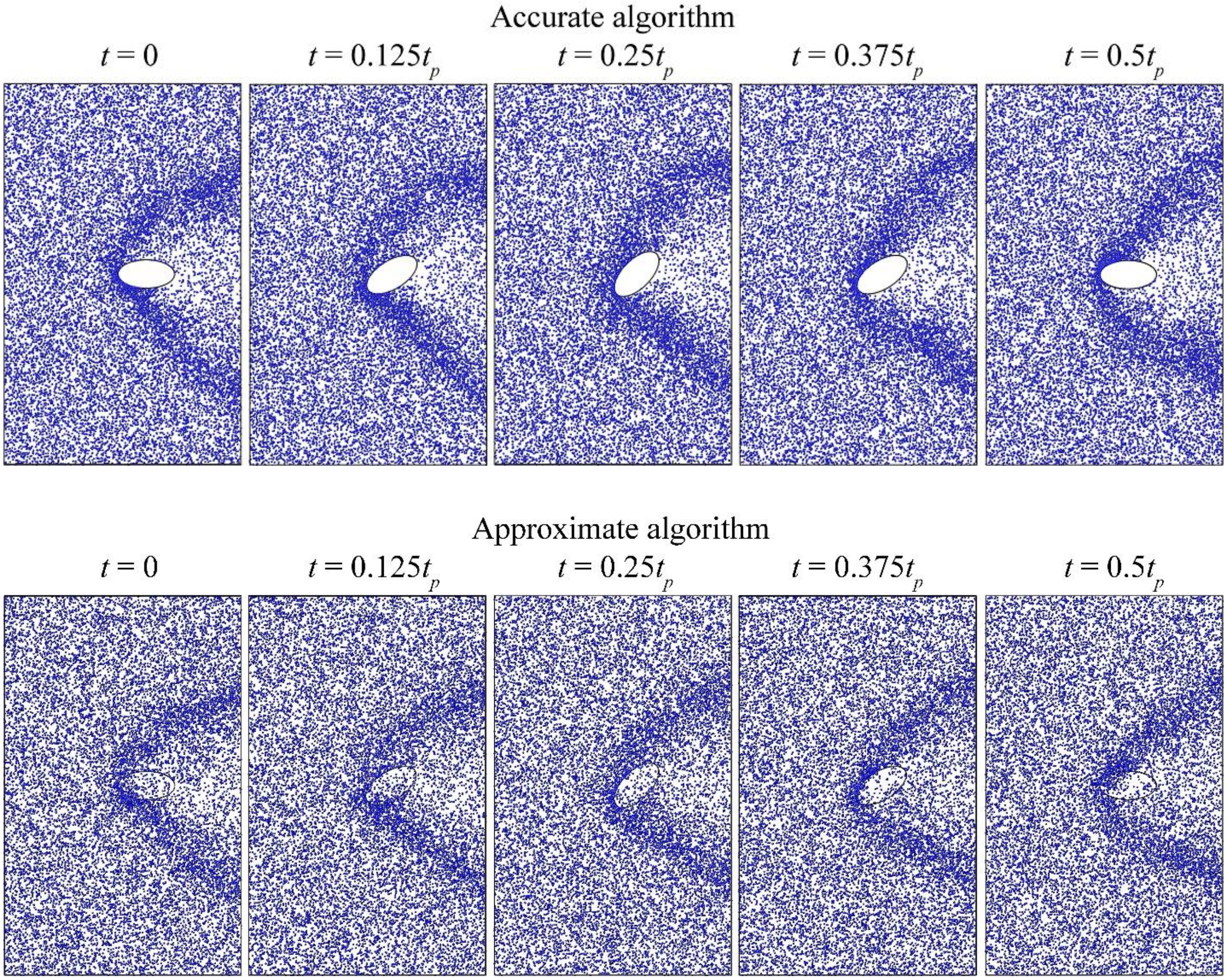

**FIG. 17.** Particle distributions obtained using the accurate (top) and approximate (bottom) algorithms for the periodically rotating ellipse ($f_p$ = 125 Hz).

In order to indicate to what extend will the computational efficiency of the accurate algorithm decrease, the time consumptions of the accurate and approximate algorithms are given and compared. With 120-core parallel computing for 10,000 time steps at $f_p$ = 31.25 Hz, the time consumptions of the two algorithms are 838.23 s and 617.65 s, respectively. It is shown that the efficiency of the accurate algorithm is 35.71 % lower than that of the approximate one. This is a fairly moderate decrease in efficiency considering that the accurate algorithm can avoid the gas leakage. In unsteady flows, the position and attitude of moving object vary with time, and therefore, the computation of cutting the surface grids from the background Cartesian grids, which is time consumed, has to be conducted in each time step. For this reason, the accurate determination of the moment and position of the particle-wall collision will not greatly increase the time consumption.

### D. Three-dimensional Apollo Command Module

The re-entry module is a typical space vehicle that travels through the rarefied atmosphere at a speed of thousands of meters per second. The Apollo command Module (ACM) is one of the most famous re-entry modules and widely adopted as a benchmark case for numerical validation. In this work, the ACM[39] is employed to validate the accurate algorithm in three dimensions.

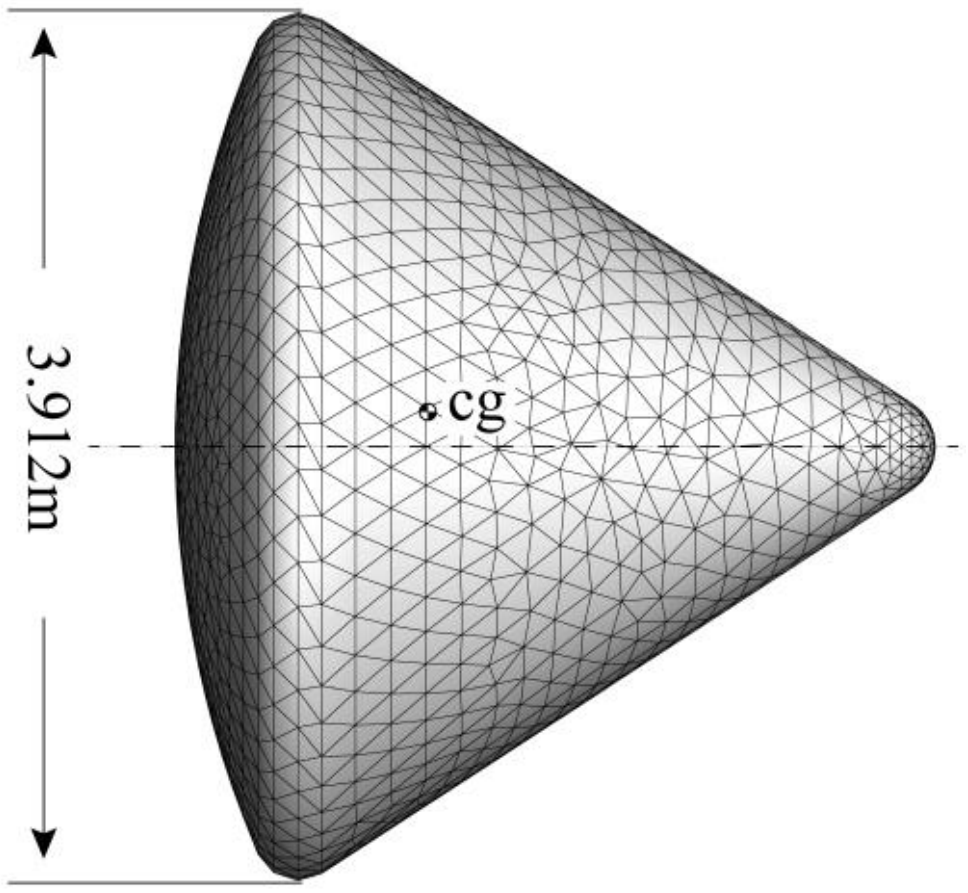

**FIG. 18.** Axisymmetric view of the ACM geometry and surface grids.

Figure 18 shows the axisymmetric view of the ACM geometry. The ACM rotates around its center of gravity (cg) in the X-Y centerplane in a Mach 10.2 flow (nitrogen) that corresponds to $Kn_\infty = 0.067$. The temperature of the incoming flow is 142.2 K. In the present simulation, the variation of the rotation angle $\theta$ also follows the sinusoidal function shown in Eq. (30), in which $f_p$ and $\theta_{\max}$ are set to 156.25 Hz and 30°, respectively.

The calculation domain is comprised of 2,766 surface grids as shown in Fig. 18 and $1.98\times10^6$ volume grids. The volume grids are multi-level Cartesian grids and the grid length scale can achieve as small as 0.03 m. The time step is $2.0\times10^{-6}$ s. The scaling factor is $6.0\times10^{13}$ and the number of particles is $1.48\times10^8$. The computational results are also temporally averaged over 100 cycles.

The surface pressure coefficient distributions in half time period predicted by the accurate algorithm are displayed in Fig. 19. The particle distributions in the X-Y centerplane, colored by $(v-v_\infty)/v_m$ with $v_m=\sqrt{3k_BT_\infty/m}$ representing the molecular mean velocity of the incoming flow, are also displayed in the figure. It is noted that both distributions at $t$ = 0 are noticeably different from that at $t = 0.5t_p$, though the ACM is in the same attitude at the two moments, indicating that the flow is highly unsteady.

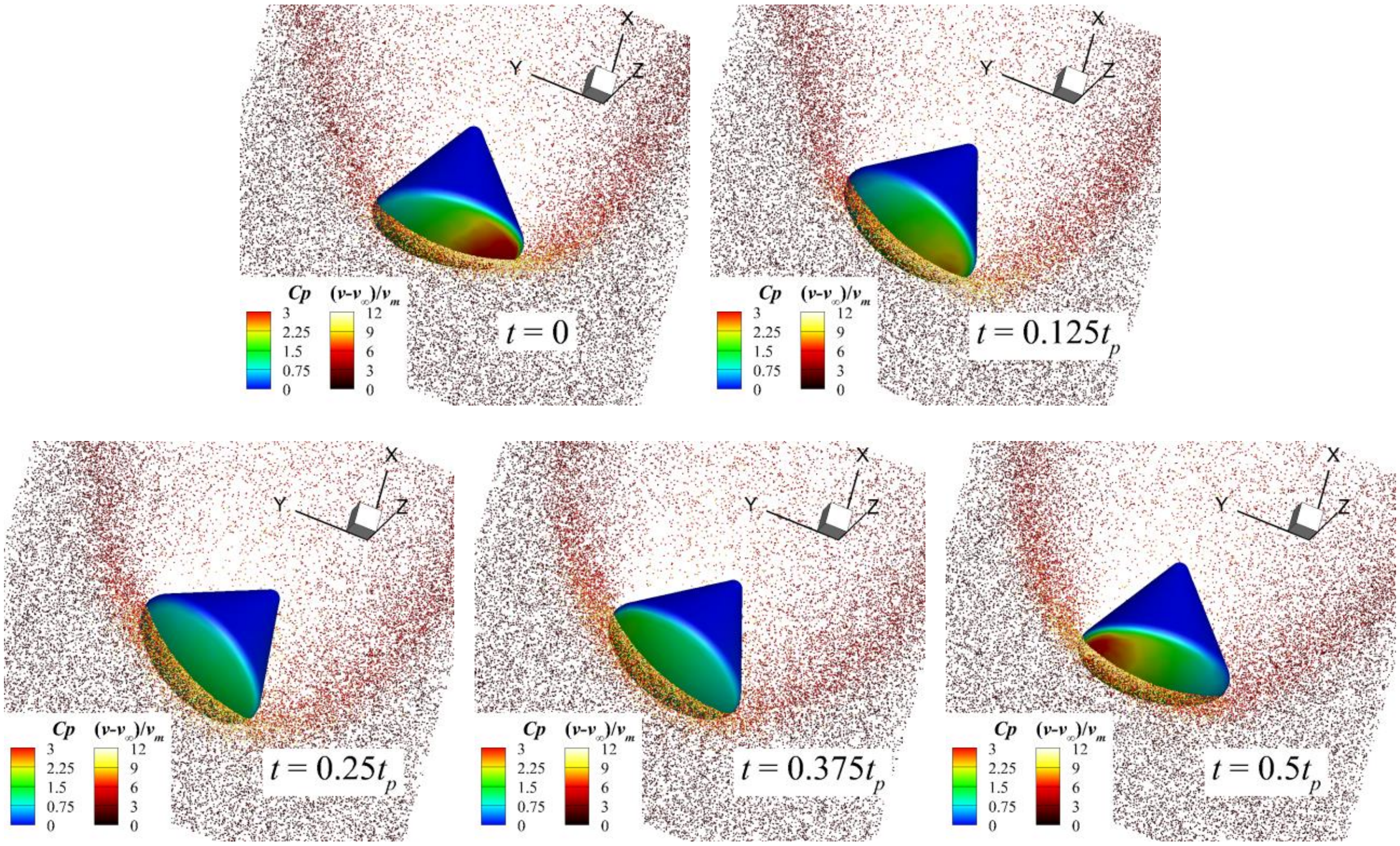

**FIG. 19.** Surface pressure coefficient distributions and particle distributions in the X-Y centerplane obtained using the accurate algorithm for the ACM.

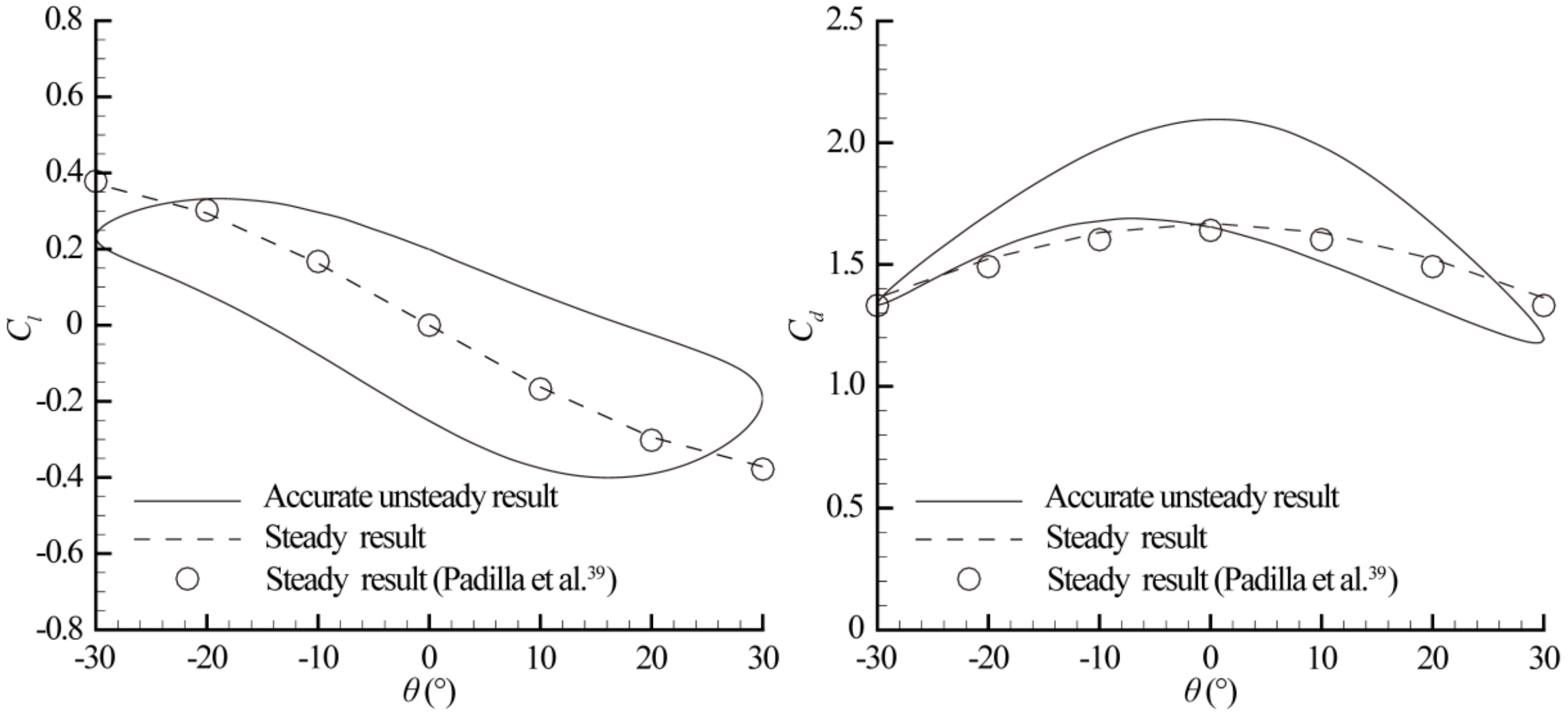

**FIG. 20.** Lift (left) and drag (right) coefficients in terms of the rotation angle for the ACM.

Figure 20 shows the variations of the lift and drag coefficients in terms of the rotation angle and compares the results of the unsteady and steady simulations. It is seen that the steady result obtained in the present work is very close to that in Padilla et al.'s wrok[39] with the use of a DSMC solver MONAGO. Besides, the unsteady result obtained using the accurate algorithm illustrates a noticeable hysteresis effect that is characterized by the hysteresis loops of the lift and drag coefficients. This result is significantly different from the steady one. It is indicated that the steady simulation may cause a failure in the predictions of lift and drag coefficients in practical unsteady problems, and thus an unsteady simulation with an appropriate algorithm as proposed in this work is essentially required. It is also seen that the hysteresis loops are asymmetric about $\theta = 0°$. This is because the ACM rotates around its center of gravity that is not on the geometric central line as shown in Fig. 18.

## VI. CONCLUSIONS

In the present work, an accurate algorithm, aimed at dealing with moving walls within the DSMC framework, is proposed and validated in rarefied flow problems. The conclusions are made in the following.

1. The accurate algorithm removes the frozen-wall assumption that is employed by the approximate algorithm, thus allowing the prediction of the particle-wall collision in a coupled manner.
2. By theoretically constructing the trajectory equation of a particle in collision with an arbitrarily shaped object of which the motion incorporates both translation and rotation, the moment and position of the particle-wall collision can be accurately determined.
3. The error analysis is conducted for the approximate algorithm. It is revealed that the error induced by this algorithm is increased with the wall velocity, indicating that the error is nonnegligible if wall moves at a high speed.
4. It is shown that the accurate algorithm is able to capture the one-dimensional normal shock wave induced by a moving piston, reproduce the equilibrium state inside a rotating square box, and characterize the unsteady rarefied flows around a periodically rotating ellipse and the ACM. The simulated results are perfectly consistent with the Maxwellian theoretical solutions. Besides, the algorithm essentially ensures particle conservation by accurately predicting the particle-wall collision and thus avoid gas leakage. The approximate algorithm, in contrast, tends to induce gas leakage since it freezes wall during the collision. Both the theoretical analysis and simulated results indicate that the gas leakage induced by the approximate algorithm gets more severe as the wall velocity is increased. It is also shown in the ACM case that the unsteady simulation with an appropriate algorithm as proposed in the present work is essentially required for reproducing the hysteresis effect in such applications.

## ACKNOWLEDGMENTS

The financial assistance provided by the National Key Research and Development Program of China (No. 2019YFB1704204) is gratefully acknowledged.

## DATA AVAILABILITY

The data that support the findings of this study are available from the corresponding author upon reasonable request.